\documentclass[twocolumn,showpacs,preprintnumbers,amsmath,amssymb]{revtex4}

\usepackage{graphicx}
\usepackage{dcolumn}
\usepackage{bm}

\newcommand{\e}{\mathrm{e}}
\newcommand{\dd}{\mathrm{d}}
\newcommand{\ft}[2]{{\textstyle\frac{#1}{#2}}}
\newcommand{\pd}{\partial}

\begin{document}


\title{Landau problem for bilayer graphene:\\ Exact results}

\author{Young-Hwan Hyun}%
\author{Yoonbai Kim}
\author{Corneliu Sochichiu}
\altaffiliation[Also at ]{University College,
Sungkyunkwan University, Suwon 440-746, Korea and Institutul de Fizic\u{a} Aplicat\u{a}, Chi\c{s}in\u{a}u}
\author{Min-Young Choi}
\affiliation{Department of Physics, BK21 Physics Research
Division, and Institute of Basic Science\\
Sungkyunkwan University, Suwon 440-746, Korea}%

\email{yhhyun, yoonbai, sochichi, mychoi22@skku.edu}

\date{\today}

\begin{abstract}
We consider graphene bilayer in a constant magnetic field of arbitrary orientation (i.e. tilted with respect to the graphene plane). In the low energy approximation to tight binding model with Peierls substitution, we find the exact spectrum of Landau levels. This spectrum is two-fold degenerate in the limit of large in-plane field, which gives rise to a new SU(2) symmetry in this limit.
\end{abstract}

\pacs{Valid PACS appear here}
\maketitle

\section{\label{sec:intro}Introduction}
There are not many examples of exactly solvable problems in physics. It is even more intriguing, when such a problem emerges in a hot topic with a rich perspective of potential applications. Graphene, one or a few layers of two-dimensional hexagonal carbon lattice of graphite
\cite{PhysRev.71.622}, attracted enormous interest in recent years due to its unusual properties, like (quasi)relativistic quasiparticle spectrum as well as prospective applications to construction of electronic devices (see \cite{2009RvMP...81..109C} for a recent review).
Among these, the bilayer graphene is believed to be one of the most rich in features at the same time accessible technologically. In particular, it exhibits a gap in the electronic spectrum, which can be tuned by external electromagnetic fields. This explains why the bilayer graphene in external fields, in particular, in a magnetic field is a subject of intensive recent study \cite{mccann:086805,Feldman2009,PhysRevLett.104.066801,Gorbar2010}.

In the present letter we consider the general Landau problem for the bilayer graphene system in a constant magnetic field having an arbitrary direction with respect to the graphene plane. We show that the energy spectrum can be found \emph{exactly} in terms of \emph{Spheroidal functions}, which are solutions to a particular case of \emph{Confluent Heun's equation} \cite{ronveaux1995}, and corresponding eigenvalues. Further analysis of the eigenvalues reveals that in the limit of strong in-plane magnetic field the
spectrum becomes two-fold degenerate, which leads to additional SU(2) symmetry.

The plan of the remainder of this note is as follows. In the next section we introduce, starting from the tight-binding Hamiltonian, the low energy Lagrangian describing the electronic wave function in bilayer graphene in external electromagnetic field. Then, we consider constant magnetic field. As a warm-up exercise, we derive the Landau level (LL) spectrum in perpendicular magnetic field. After this we consider the generic case of tilted magnetic field, and find the eigenvalue spectrum. Finally, we analyze the spectrum and reveal an asymptotic two-fold degeneracy as well as compare it to the purely in-plane magnetic field result. The technical details behind the results reported here will appear in an accompanying paper \cite{our-big-paper}.

\section{Bilayer graphene in constant magnetic field}
\begin{figure}
\includegraphics{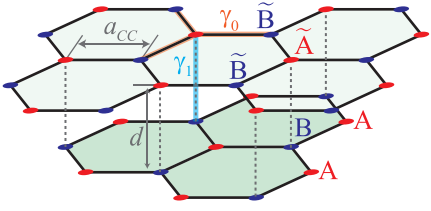}
\caption{\small Schematic description of the graphene bilayer lattice with Bernal stacking, $a_{cc}\approx 1.42$\AA{} and $d\approx 3.35$\AA. The dotted lines represent the interlayer tunneling. We use tildes for the parameters referred to the upper layer and no tildes for the lower layer ones.}
\label{fig01}
\end{figure}

According to the tight-binding model the bilayer graphene with Bernal stacking can be described by the following Hamiltonian,
\begin{multline}
H = -\gamma_0\sum_{{\bf n},a,\sigma}
\left({\tilde a_{{\bf n},\sigma}^{\dagger}}
{\tilde b_{{\bf n}+\boldsymbol{\delta}_{a},\sigma}}
+{\tilde b_{{\bf n}+\boldsymbol{\delta}_{a},\sigma}^{\dagger}}
{\tilde a_{{\bf n},\sigma}}
\right)\\
-\gamma_0\sum_{{\bf n},a,\sigma}
\left({b_{{\bf n},\sigma}^{\dagger}}
{a_{{\bf n}-\boldsymbol{\delta}_{a},\sigma}}
+{a_{{\bf n}-\boldsymbol{\delta}_{a},\sigma}^{\dagger}}
{b_{{\bf n},\sigma}}
\right)\cr
-\gamma_1\sum_{{\bf n},\sigma} \left(\tilde a_{{\bf n},\sigma}^\dag
b_{{\bf n},\sigma} +b_{{\bf n},\sigma}^\dag \tilde a_{{\bf n},\sigma}\right),
\label{H0}
\end{multline}
where $\tilde{a},\tilde{b}$ and their conjugate $\tilde{a}^\dag,\tilde{b}^\dag$ are, respectively annihilation and creation operators for $\tilde A$ and $\tilde B$ sites of the upper layer, while lower layer operators carry no tildes. The indices $\mathbf{n}$ run through $\tilde A/B$ lattice sites, $\boldsymbol{\delta}_{a}$, $a=1,2,3$ are the vectors connecting this site to the nearest neighbors on the upper layer, $-\boldsymbol{\delta}_{a}$ does the same for the lower layer and $\sigma=\pm$ is the electron's spin. Parameter $\gamma_{0}$ is the nearest neighbor hopping amplitude, and $\gamma_{1}$ is the leading interlayer tunneling amplitude (see fig. \ref{fig01}). The experimental values of the couplings
 found e.g. in~\cite{PhysRevB.76.201401} are,
\begin{align}\label{tamu}
\gamma_{0} \approx 2.9\,\mbox{eV}, \qquad \gamma_{1}\approx 0.3\,\mbox{eV}.
\end{align}
Electromagnetic interaction is turned on through the Peierls substitution, i.e. any two operators located at different sites, e.g. ${\bf n}$ and ${\bf n}+\boldsymbol{\delta}$ are connected by a phase factor representing the parallel transport in electromagnetic field,
\begin{align}\label{Und}
  U_{{\bf n},{{\boldsymbol{\delta}}}}\equiv U_{{\bf n}+{\boldsymbol\delta},{\bf n}}
  \approx\exp \left( -\frac{i e}{\hbar c}\int_{\bf n}^{{\bf n}+{\boldsymbol\delta}}
  {\bf A}\cdot\dd{\bf x} \right)~.
\end{align}

At low energy the quasiparticle is described by an eight component wave function $\Psi({\bf x},t)$: There are two components for each Dirac valley ${\pm\bf K}$ as well as two values of the spin index.
Although a subject to quadratic dispersion relations, the quasiparticle is a chiral particle.

In the case of constant magnetic field the effective description can be casted into the following Lagrangian,
\begin{align}
{\cal L}_{\rm eff} = \Psi^\dag \left[i\hbar\frac{\partial}{\partial t} - \begin{pmatrix}
{\cal H}_{({\bf K})} & 0  \\ 0 & {\cal H}_{(-{\bf K})} \end{pmatrix}
\right]\Psi~,\label{Lemtot}
\end{align}
where the chiral `Hamiltonians' ${\cal H}_{({\pm\bf K})}$ correspond to different Dirac points ${\pm\bf K}$, where ${\bf K}=(4\pi/3\sqrt{3}a_{cc},0)$ \footnote{We are following the convention in which the nearest neighbors are connected by the vectors: $\boldsymbol{\delta}_1=a_{cc}(0,1)$, $\boldsymbol{\delta}_2=a_{cc}(\sqrt{3}/2,-1/2)$ and $\boldsymbol{\delta}_3=a_{cc}(-\sqrt{3}/2,-1/2)$.}. The operators ${\cal H}_{({\pm\bf K})}$  are most conveniently expressed using the complex coordinate for the graphene plane, $w=\ft1{\sqrt{2}}(x+i y)$. Thus ${\cal H}_{({\bf K})}$ is given by
%
\begin{equation}\label{hk}
{\cal H}_{({\bf K})} =-\frac{\hbar^2 v_{\text{\tiny F}}^2}{\gamma_1} \left(\begin{array}{cc} 0& {D}
 e^{+i{\Phi}} {D} \\
  \bar{D}
   e^{-i{\Phi}}
 \bar{D} &  0  \end{array}\right)
,
\end{equation}
%
while the operator ${\cal H}_{(-{\bf K})}$ is obtained from ${\cal H}_{({\bf K})}$ by interchange of $D$ and $\bar{D}$.
In eq. \eqref{hk} we neglected the Zeeman term, the only effect of which is a shift of energy eigenvalues by
$\pm (\hbar e/2m_{\rm e} c) B$. The covariant derivative $D$ and its conjugate $\bar{D}$ are defined, respectively, as $D \equiv \partial_w +\frac{ie}{\hbar c}A_w$, $\bar{D} \equiv \partial_{\bar{w}} +\frac{ie}{\hbar c}A_{\bar{w}}$,
$A_{w}$ and $A_{\bar{w}}$ being the complex components of the vector potential. Peierls' phase
\begin{equation}
\Phi=\frac{e}{\hbar c}\int_0^dA_z\dd z,
\end{equation}
corresponds to the parallel transport between the layers.

In the case of a constant magnetic field we can fix the gauge such that electromagnetic potential $A$, is given by
\begin{equation}
A_w =A_{\bar w}^*= -\frac{i}{2}B_\perp \bar w,\quad A_z = \frac{i}{\sqrt2}B_{\parallel} (\eta\bar w
-\bar\eta w).
\end{equation}
Here $B_\perp$ and $B_{\parallel}\eta$ are, respectively, perpendicular and in-plane components of the magnetic field, the phase factor $\eta$ gives the direction of the in-plane projection of magnetic field and can be absorbed into redefinition of $w$. We restrict our analysis to a single Dirac point ${\bf K}$ and, respectively, drop the subscript.

If the in-plane component $B_{\parallel}$ vanishes, i.e. the magnetic field is perpendicular to the graphene plane, $B_\perp=B$, the Hamiltonian \eqref{hk} reduces to,
\begin{equation}\label{h-perp}
{\cal H} =-\ft{\hbar^2}{2m} \left(\begin{array}{cc} 0& {D}^2
  \\
  \bar{D}^2
    &  0  \end{array}\right),
\end{equation}
where $m=\gamma_1/2v_{\text{\tiny F}}^2\approx 0.030\times m_{\rm e}$ is effective mass of the fermion.

The operator ${\cal H}$ can be readily diagonalized by standard techniques. Indeed, introduce the oscillator raising/lowering operators $\alpha$ and $\alpha^\dag$ with standard commutation relations according to
\begin{equation}
\alpha^\dag \equiv -{\sqrt{\frac{\hbar c}{eB}}}D,\quad
\alpha \equiv {\sqrt{\frac{\hbar c}{eB}}}\bar{D}.
\end{equation}
Then the problem is equivalent to the eigenvalue problem
\begin{equation}\label{evprob}
  \begin{pmatrix}
    0 & \alpha^\dag{}^2 \\
    \alpha^2 & 0
  \end{pmatrix}
  \begin{pmatrix}
    \psi_1\\
    \psi_2
  \end{pmatrix}
  =\varepsilon
  \begin{pmatrix}
    \psi_1\\
    \psi_2
  \end{pmatrix},
\end{equation}
where $\varepsilon$ corresponds to the energy eigenvalue $E$ as $E=-(\hbar e B/mc)\varepsilon$.

The solution to the eigenvalue problem \eqref{evprob} is given by
\begin{equation}
  E_n=\pm\hbar\left(\frac{e B}{mc}\right)\sqrt{n(n-1)},
\end{equation}
parameterized by $n=0,1,2,\dots$ and the sign $\pm$. Let us note that due to chirality the Hamiltonian is not sign definite.

Now let us turn on the in-plane magnetic field and consider the modifications in the eigenvalue problem. The Hamiltonian now can be written as
\begin{equation}\label{Hinplane}
   {\cal H} = -\frac{\hbar eB_\perp}{mc}
   \begin{pmatrix}
   0 & \alpha^\dag e^{-\nu(\bar w- w)}
\alpha^\dag \\ \alpha e^{
\nu(\bar w- w)}\alpha & 0
 \end{pmatrix}
  ,
\end{equation}
where $\nu=(ed B_{\parallel}/\hbar c \sqrt2)$. One can easily find the zero modes of this operator, so let us concentrate on the nonzero eigenvalues. In this case the eigenvalue problem for Hamiltonian \eqref{Hinplane} can be reduced to the following one component problem,
\begin{equation}
  \label{originalEq}
\alpha^\dag(\alpha^\dag +\beta)(\alpha+\beta)\alpha \psi_1 = \lambda \psi_1,
\end{equation}
with $\beta= \nu {\sqrt{\frac{\hbar c}{eB_\perp}}}$ and $\lambda=\varepsilon^2$. The remaining component $\psi_2$ is expressed through,
\begin{equation}
  \psi_2=\lambda^{-1}\e^{\nu(\bar w-w)}(\alpha+\beta)\alpha \psi_1.
\end{equation}
The one-component problem \eqref{originalEq} is not as simple as in the purely perpendicular case, so let us consider it more closely. Using the antiholomorphic representation:
\begin{equation}
  \alpha^\dag \mapsto \bar z, \quad \alpha\mapsto \frac{\dd}{\dd \bar z},
\end{equation}
the eigenvalue problem is reduced to the second order differential equation
\begin{equation}\label{eqMain}
x(x+\beta) (u'' +\beta u') -\lambda u = 0,
\end{equation}
which we can solve on the real line and then analytically continue.
The finite norm condition for eigenstate $\psi$ implies that function $u$ should be a regular function in the interval $x\in (-\infty,\infty)$ with a moderate growth (at most exponential) as $x\to\pm\infty$.

The equation \eqref{eqMain} is a particular case of confluent Heun's or spheroidal equation. By a change of independent variable $x\mapsto (\beta/2) (x-1)$ and then of the dependent one $u(x)= \sqrt{x^2-1}\exp({-\frac{\beta^2}{4}x})v(x)$ it reduces to one of the standard forms of the spheroidal equation. The solutions satisfying physical conditions are given by the \emph{angular oblate spheroidal functions},
\begin{equation}
  u(x)=\frac{c_n(\beta)}{\sqrt{x(x+\beta)}}\e^{\frac{\beta}{2}x}\Xi_{1,0,n-1}
  \left(\ft{\beta^2}{4},0,\ft{2x}{\beta}+1\right),
\end{equation}
where $c_n(\beta)$ is the normalization constant. These functions represent Landau eigenstates.
The corresponding Landau levels are given by the spheroidal eigenvalues~\footnote{We use the notations of \cite{ronveaux1995} in which our spheroidal equation is a particular form of Confluent Heun's equation (CHE):
$
  \frac{\dd}{\dd z}(z^2-1)v'+\left(-p^2(z^2-1)+2 p b z
-\lambda- \frac{m^2+s^2+2msz}{z^2-1}\right)v=0
$, with $s=0$. The eigenvalues of CHE are $\lambda^{(a)}_{m,s,n}(p,b)$. }
\begin{equation}
  E_n^2=\left(\frac{\hbar e B_\perp}{mc}\right)^2
  \lambda^{(a)}_{1,0,n-1}\left(\frac{\beta^2}{4},0\right),
\end{equation}
where the superscript $(a)$ stands for `angular'. Recall, that $\beta=\sqrt{{ed^2}/{2\hbar c}}\ ({B_{\parallel}}/{\sqrt{B_\perp}})$.

Spheroidal equations appear in many situations, like molecular hydrogen ion eigenvalue problem, radio antenna description as well as geophysical applications, therefore spheroidal functions and respective eigenvalues are relatively well studied \cite{ronveaux1995}. They are also implemented as standard libraries in \texttt{Mathematica} \cite{wolfram}.

\begin{figure}
\includegraphics[scale=0.8]{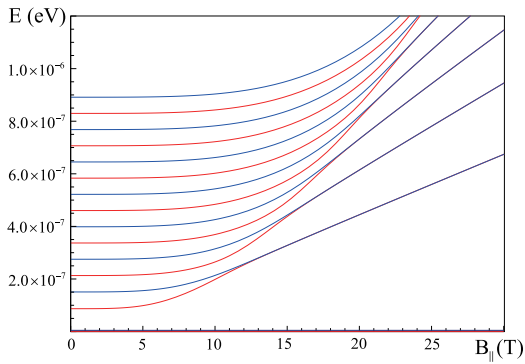}
\caption{\small A plot of Energy eigenvalues (eV) as a function of in-plane magnetic field (T) for the fixed value of perpendicular field $B_\perp=10^{-4}$T. Red lines represent even energy levels, while blue lines the odd ones. One can clearly see convergence of even and odd neighbor eigenvalues. Zeroth and first states are the zero modes of the Hamiltonian, therefore, they are degenerate at any value of field.}
\label{fig:eigen}
\end{figure}

A typical behavior of energy eigenvalues is depicted in Fig. \ref{fig:eigen}. From it one can see, that for $\beta=0$ the eigenvalue spectrum is reproducing that of strictly perpendicular magnetic field problem. These eigenvalues are quite stable against small in-plane field perturbations. Indeed, small $\beta$ expansion for the spheroidal eigenvalues gives,
\begin{widetext}
\begin{equation}
  \left.\left(\frac{mc}{\hbar e B_\perp}\right)^2E_n^2
  \right|_{\beta\to 0}=
     n(n-1)\\
     +
     \frac{n(n-1)}{32n(n-1)-24}\beta^4
     +\frac{n(n-1)(n(n-1)(4n(n-1)-39)+63)}{128(2n+3)(2n+1)^3(2n-3)^3(2n-5)}
     \beta^8
     +O(\beta^{10}),
\end{equation}
\end{widetext}
so, the leading correction is of the order $\sim B_\parallel^4/B_\perp^2$, which means that for a reasonably small tilt the LL spectrum is very close to the purely perpendicular magnetic field values.

In the opposite case, when the magnetic field is almost parallel to the graphene plane, we can use the large $\beta$ asymptotic expansion of the eigenvalues \cite{russian_book},
\begin{multline}\label{largeB}
  \left.\left(\frac{mc}{\hbar e B_\perp}\right)^2E_n^2
  \right|_{\beta\to \infty}=\\
  \left[\ft{n}{2}\right] \beta^2-2\left[\ft{n}{2}\right]^2
  -\left[\ft{n}{2}\right]^3\ft{4}{\beta^2}+\dots,
\end{multline}
where the square bracket denotes the integer part. The presence of this function implies that there is an asymptotic degeneracy between the each even level and the following odd one. The difference between these levels is exponentially vanishing:
\begin{equation}
  \left(\frac{mc}{\hbar e B_\perp}\right)^2  (E_{2k+1}^2-E_{2k}^2)=
  \frac{2\beta^{4k+2} }{(k-1)!\,k!} \e^{-\frac{\beta^2}{2}}+\dots,
\end{equation}
where $k=0,1,2,\dots$
This degeneracy can be interpreted as emergence of a new SU(2) symmetry in the nearly in-plane limit of the orientation of magnetic field. This asymptotic symmetry is mixing the neighbor states with numbers $2k$ and $2k+1$.

Although, mathematically even and odd eigenvalues are quickly converging to each other, in practice, the effect can be caught only at very large values of the magnetic field and/or very precise in-plane orientation.  Indeed, for magnetic field measured in tesla, $\beta\sim 10^{-3}B_\parallel/\sqrt{B_\perp}$. Assuming that degeneracy for the lowest levels occurs at the values $\beta\sim 10$, we have $\delta\theta\sim 10^{-7} B$, where $\delta\theta$ is the angle deviation from the parallel direction. Currently available magnetic fields are of the order $\sim 10^2$T, which means that to capture a measurable effect the magnetic field should be parallel to the graphene plane with the accuracy $\sim 10^{-5}$rad.

More generally, the asymptotic expansion formula \eqref{largeB} predicts that in the limit in which in-plane component is much larger than the perpendicular one the energy eigenvalues scale like $E_n\sim B_\parallel\sqrt{B_\perp}$ or in terms of magnitude and angle as $\sim B^{3/2}\sqrt{|\theta|}$. Thus, the energy eigenvalues in such a magnetic field exhibit a hybrid behavior revealing both relativistic and non relativistic patterns. Thus, if we fix the (large) value of the in-plane $B_\parallel$ the spectrum scales according to the relativistic rule: $E\sim\sqrt{B_\perp}$.

It is instructive to compare the above limit with the exact solution in parallel magnetic field, which we can obtain after setting $B_\perp=0$.

In the parallel magnetic field the Dirac point Hamiltonian \eqref{hk} becomes,
\begin{equation}\label{hkparallel}
 {\cal H} =-
 \ft{\hbar^2}{2m}
 \begin{pmatrix} 0& {\pd}
 e^{-\nu(\bar{w}-w)} {\pd} \\
  \bar{\pd}
   e^{\nu(\bar{w}-w)}
 \bar{\pd} &  0
\end{pmatrix}
.
\end{equation}
This operator can be easily diagonalized using the plane wave basis $u_{\bar{k},k}\sim\e^{-(k\bar{w}-\bar{k}w)}$, which gives a continuous spectrum of eigenvalues parameterized by the wave number $k$:
\begin{equation}\label{parallel-disp}
   E_{\bar{k},k}^2=\ft{\hbar^4}{4m^2}|k|^2|\nu-k|^2
\end{equation}
where $\nu$ is defined below the equation \eqref{Hinplane}. In addition to the invariance with respect to inversions about $x$-axis, the spectrum is left invariant by the transformation $k\mapsto \nu-k$. In particular, there are two zeroes of the dispersion relation \eqref{parallel-disp}: at $k=0$ and $k=\nu$. The group of symmetry transformations mixing the waves $u_{\bar{k},k}$ and $u_{\nu-\bar{k},\nu-k}$ is SU(2).

In the low energy regime (in this case, when $E\ll\ft{\hbar^2 v_{\text{\tiny F}}^2}{\gamma_1}\nu^2$) the energy can be expanded around the zeroes of dispersion relations: $k=0$ and $k=\nu$. This yields two massless Dirac fermions separated by momentum shift $\nu$ as discussed in \cite{Pershoguba2010}. The quantum numbers counting these `vacua' are related to the SU(2) group which describes the symmetry of the spectrum. In this limit the whole electronic system is described by a sixteen component wave function.

It is clear, that this SU(2) symmetry can be identified with the asymptotic SU(2) in the large $\beta$ limit of the tilted magnetic field system. The easiest way to convince ourselves in this is to consider the zero eigenvalue levels of the tilted problem. Unlike the higher levels, they can be expressed in terms of elementary functions. There is a simple relation between the zeroth eigenvalue eigenfunctions $\psi_0(\bar{w},w)$ and $\psi_1(\bar{w},w)=\e^{\beta\alpha^{\dag}}\psi_0(\bar{w},w)=\e^{-\nu\frac{\hbar c}{e B_\perp}D}\psi_0(\bar{w},w)$. This relation implies that, in the parallel field limit $B_\perp\to 0$, the Fourier modes are related up to a constant phase factor by the transformation: $k\mapsto \nu-k$.

\begin{acknowledgments}
The authors benefited from useful discussion with Philip Kim.

This work was supported by NRF research grant nr. 2010-0007637.
\end{acknowledgments}


\end{document}